# Fractal Behavior of Flow of an Inhomogeneous Fluid Over a Smooth Inclined Surface

N. Maleki-Jirsaraei, B. Ghane-Motlagh, S. Baradaran, E. Shekarian, S. Rouhani[*]

*Complex Systems Laboratory, Physics Department, Azzahra University
Tehran, Iran*
*\*Physics Department, Sharif University of Technology*
*&*
*School of Intelligent Systems (S.I.S.)
Institute for Studies in Theoretical Physics and Mathematics (I.P.M.)
Tehran, Iran*

**Abstract:**

Patterns formed by the flow of an inhomogeneous fluid (suspension) over a smooth inclined surface were studied. It was observed that for inclination angle larger than a threshold (10-12)°, global fractal patterns are formed. The fractal dimensions of these patterns were measured $df = 1.35 - 1.45$ which corresponds to that observed for the flow of water over an inhomogeneous surface, implying that this system is within the same universality class. Except that here the disorder is present in the fluid and is transferred to the surface. This observation is in agreement with the result of the theories of nonlinear flow of fluid in disordered media.

## 1. Introduction:

The flow of fluid in a random media as an instance of collective nonlinear transport with strong disorder [1-3] has attracted much attention. Macroscopic transport occurs only when the driving force exceeds a threshold magnitude. Near the threshold, there is some evidence of critical behavior, including diverging correlation length. Many interesting effects may be expected such as scaling relations between the critical exponent above and below the threshold etc. Examples of this behavior can be seen in the patterns formed on the windowpane during rain [1]. In all these examples, the environment is strongly disordered but through this environment, a simple body (flux line in the superconductors [2]), a homogeneous fluid (rain on a dirty rough surface) flows [1] or invasion of a porous media by a non-wetting fluid [3]. In contrast, to these examples the system concerning us is a system in which disorder is present within the moving fluid and the environment is homogeneous. A suspension made from yogurt and water, which flows down over an inclined glass surface. The surface is clean and the fluid does not erode it.

The length of the fractal structure of downward flow over the inclined surface is not infinite (global) if the angle of inclination is below a threshold. Above the threshold angle there exists one global structure. We obtained this angle to be between (10-12)°. For above the threshold there are more than one global fractal structures. Here we report the scaling properties of the system. Our observations are in good agreement with the theoretical results of the flow of fluid over rough surfaces [4,5].

## 2. Experimental Procedures:

The experiments were carried out with a suspension of yogurt and water in a controlled proportion. The ratio of mixture was kept fixed during all the experiments. The volume ration of yogurt to water was 100 to 150 the main advantage of this mixture is its cheapness and availability. An analysis graph of particle size is shown in figure 1 we observe that the particle size is sharply defined and is around 50$\mu m$. Specifications of this suspension obtained through measurements by particle size analyzer are given in Table 1. The viscosity of this suspension was measured to be 1.177 *m Pa. s.*

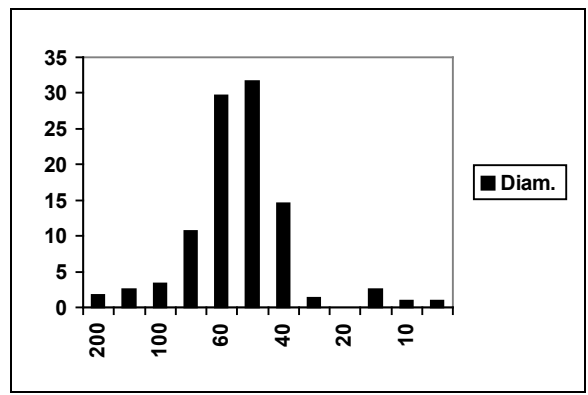

Figure 1: Particle Size Distribution

Table1: Particle Specifications.

| Median Diameter (μ m) | Modal Diameter (μ m) | Surface Area (m²/g) | Particle Density (g/cm³) |
|---|---|---|---|
| 49.37 | 48.28 | 1.14 | 1.0381 |

Then the fluid was poured down an inclined surface. The precipitation patterns, which were left by the fluid, have obvious fractal features. These patterns were registered and digitized by applying a scanner and a computer. An example of the digitized pattern is shown in figure 2. The procedure was repeated many times and a threshold angle was obtained to be (10-12)°. The length scale of patterns above and below the threshold is obviously different.

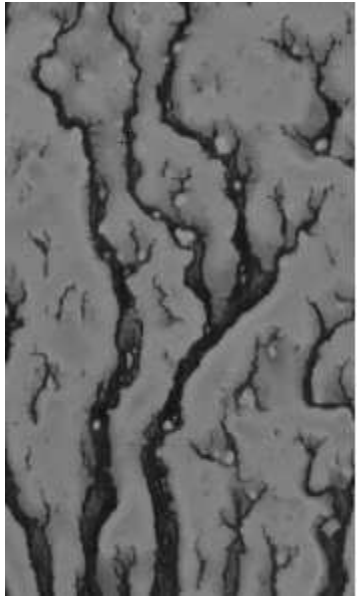

Figure 2: Digitized pattern of the flow of the yogurt-water suspension over an inclined surface

## 3. Image Analyzing Procedure:

After importing a digitized pattern, the fractal dimension is measured using various methods. For calculating the fractal dimension we used three different methods.

**Method 1:** The mass of a particular pattern, $m$, is related to its downward path of flows

$$m(\ell) \sim \ell^{df} \qquad (1)$$

In this method several patterns are formed and each of them is shown by a curve from which the fractal dimension is the calculated. The average of all curves yields the exponent $df$.

Figure 3 shows one of these graphs and its slope. In this method $df$ is obtained to be in the range 1.35-1.45.

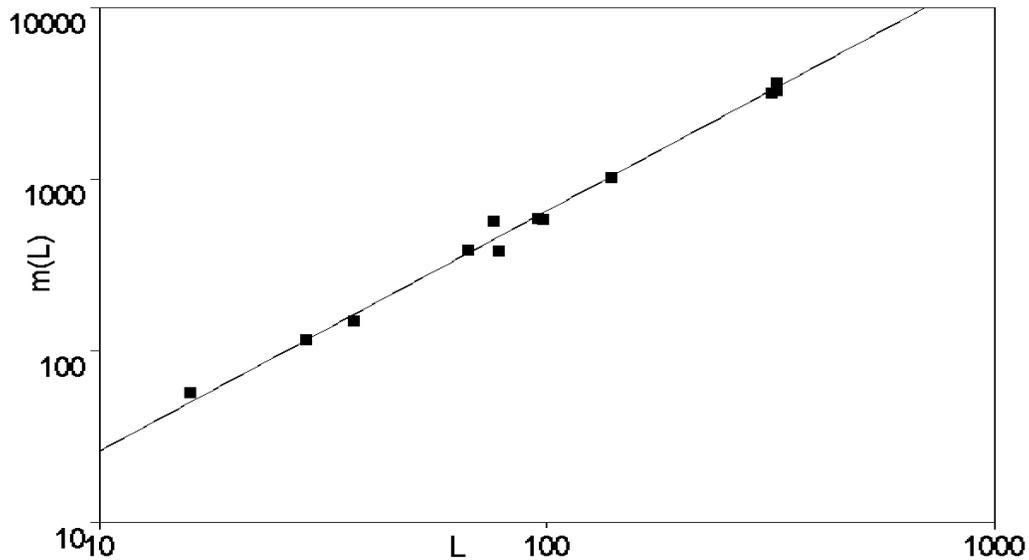

Figure 3: The fractal dimension, through scaling of mass with length, for sub branches

**Model 2:** The total mass of flows, $M$, scales with the total length as:

$$M(L) \sim L^{df} \qquad (2)$$

In this method each stream in the pattern is represented by a point instead of a curve. After drawing a log-log diagram (Figure 4), these points form a straight line whose slope is the fractal dimension of patterns. In this method the obtained fractal dimension was $df = 1.40 \pm 0.5$.

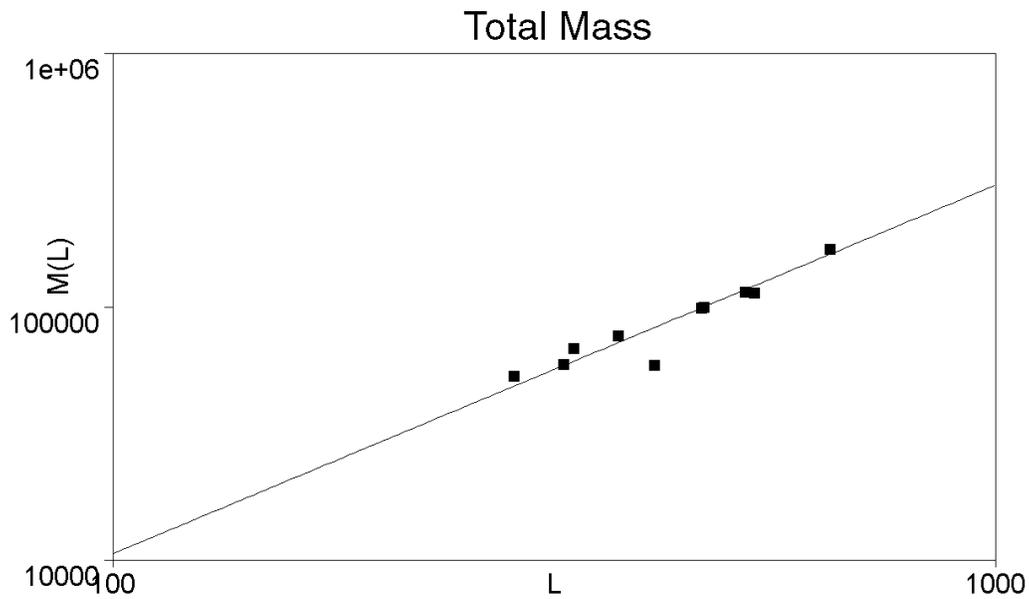

Figure 4: The fractal dimension, through scaling of mass with length

**Model 3:** The last method, which is studied, is the correlation function method [1,6], which is defined as follows

$$C(r) \equiv N^{-1} \sum_{\rho_0} \rho(r)\rho(r_0 + r) \qquad (3)$$

Here $\rho$ is the local density and if the point of location $r$ belongs to the structure $\rho(r) = 1$ and if it doesn't belong to it then $\rho(r) = 0$, $N$ is the total number of points that are used for calculating the correlation function. The function $C(r)$ represents the expected value that the two points belong to the pattern. In isotropic fractal pattern we expect $C(r)$ not depend on direction, but only the distance therefore $C(r) = C(r)$.

In the present case the horizontal and vertical directions clearly differ, therefore we expect that $r$ has to be parallel to the direction of downward.
The correlation function satisfies an exponential function in the form

$$C(r) \sim r^{-\alpha} \qquad (4)$$

The mass of stream $m(\ell)$ can be expressed in terms of the correlation function from which it follows that fractal dimension is [6]

$$df = 2 - \alpha \qquad (5)$$

We observe in figure 5 a plot of $C(r)$ versus $r$ for the downward flow:

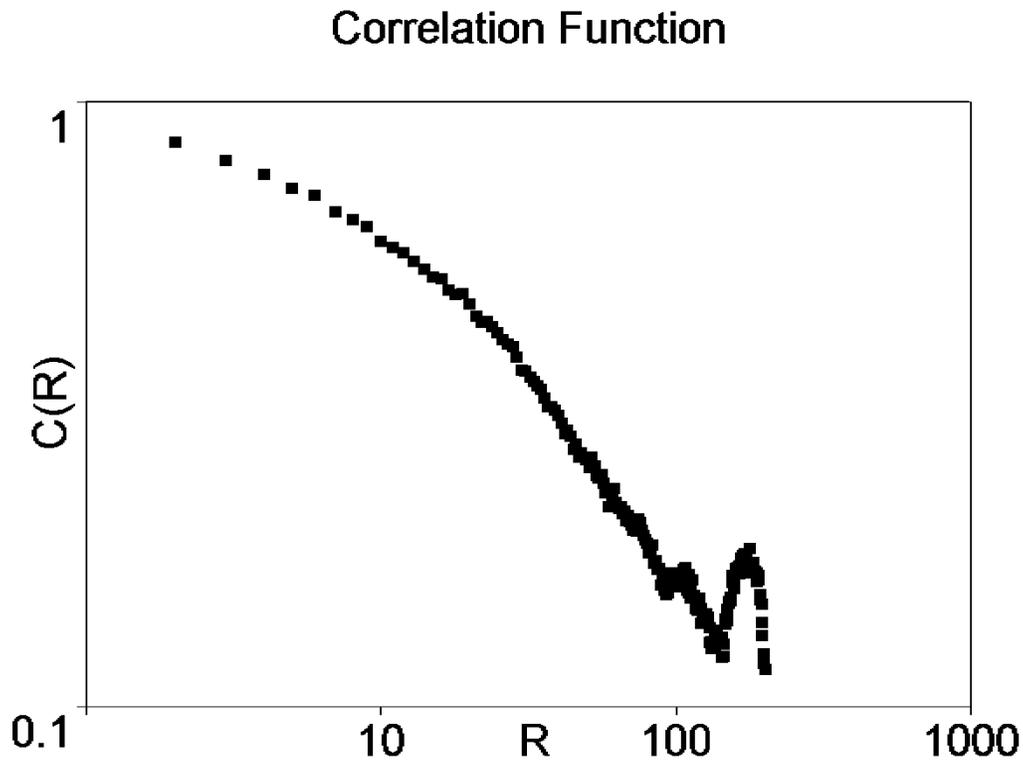

Figure 5: The correlation function for the downward direction

Finite size effects cause $C(r)$ to have a power low behavior only in its mid section. Figure 5 shows the log curves, by picking its linear portion, which is evidently different, and finding the slope of that line, we find $\alpha$ and from it the fractal dimension. Figure 6 shows the linear curve and its slope.

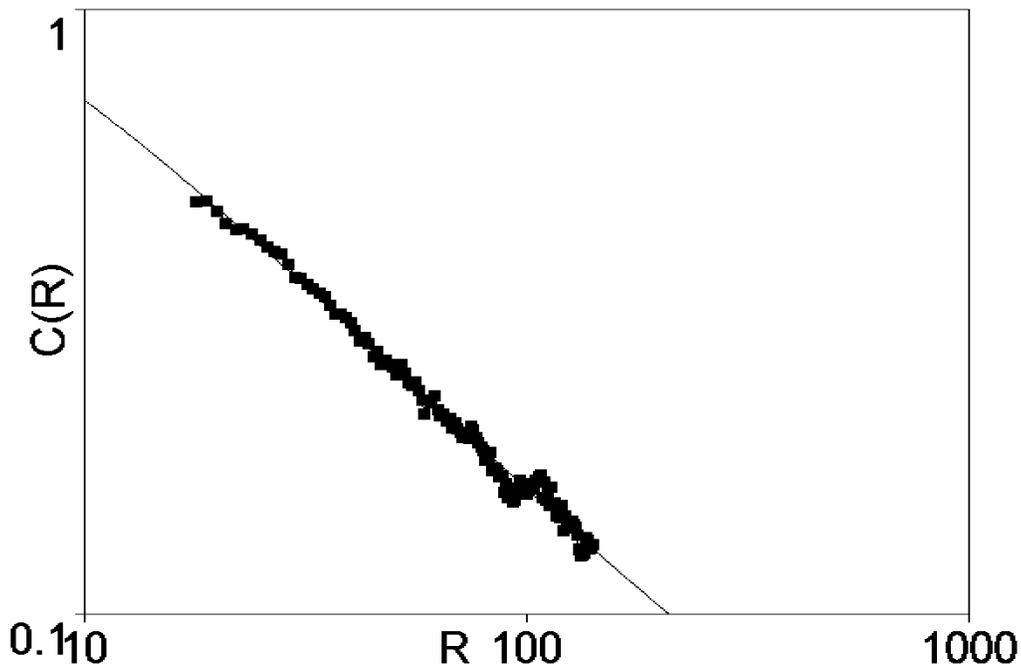

Figure 6: Fractal dimension through correlation function

**4. Conclusion:**

Narayan and Fisher [4] proposed a model for nonlinear behavior of the flow of the fluid over a rough random surface. In this model it is assumed that the disorder is strong enough to break into several channels. As the surfaces' of incline is slowly increased, the fluid collects into lakes with a certain depth, which when saturated fill the adjacent ones underneath, forming clusters of lakes. If the angle of incline is small, then the water doesn't fill the lake completely when the angle of incline is larger the length of clusters increases reaching an infinite value at a critical threshold angle. Below this limit it is possible to see many isolated clusters that are totally disconnected from the flow. At threshold, which is equivalent of the defining transition, there exists at least one flowing river from top to bottom (e.g. at least one cluster whose correlation length is infinite). The system under study is different from the model of Narayan and Fisher, because the disorder is in the fluid not in the surface in spite of this difference, the measured fractal dimension is in agreement with the predictions of Narayan and Fisher model [4] and experimental observations [1] and apparently the studied system is in the same class of universality.

Mean field theory predicts the value $\frac{4}{3}$ for the fractal dimension of this kind of grids [4] while our measured quantity $(1.40 \pm 0.5)$ in approximately equal with this, it is completely different with the fractal dimension for the river networks. In the experiment which studied the movement of water over an inclined dirty surface, the fractal dimension was measured $1.37 \pm 0.05$ which is in good agreement with our measurement.

**Acknowledgements:**
We would like to thank O. Moshtagh-Askari for helping us with the programming.